%% file: main.tex
\documentclass{article}
\RequirePackage{fix-cm}
\usepackage{times}
\usepackage[authoryear,round]{natbib}
\setcitestyle{citesep={;},aysep={,},yysep={;}}
\usepackage{fancyhdr}
\renewcommand{\headrulewidth}{0pt}
\usepackage[utf8]{inputenc}
\usepackage[T1]{fontenc}
\usepackage{amsmath,amssymb,amsfonts,booktabs,graphicx,microtype,multirow,xspace}
\usepackage[hidelinks]{hyperref}
\usepackage{url,placeins}
\newcommand{\tablecaptiongap}{\setlength{\belowcaptionskip}{11pt}}
\newcommand{\method}{\textnormal{\textsc{AnchorFlow}}\xspace}

\title{AnchorFlow: Learning Anchor Placement for Faithful and Editable SVG Reconstruction}
\author{%
\begin{tabular}{c}
Mengnan Jiang$^{1,2}$ \quad Christian Franke$^{1}$ \quad Michele Franco Adesso$^{1}$\\[4pt]
Antonio Haas$^{1}$ \quad Grace Li Zhang$^{2}$\\[8pt]
\small $^{1}$Mercedes-Benz AG \quad $^{2}$Technical University of Darmstadt\\[4pt]
{\scriptsize\texttt{mengnan.jiang@mercedes-benz.com}\quad\texttt{christian.franke@mercedes-benz.com}}\\
{\scriptsize\texttt{michele\_franco.adesso@mercedes-benz.com}\quad\texttt{antonio.haas@mercedes-benz.com}}\\
{\scriptsize\texttt{grace.zhang@tu-darmstadt.de}}
\end{tabular}}
\date{}
\begin{document}
\maketitle\thispagestyle{fancy}
\begin{abstract}
Raster-to-SVG reconstruction requires faithful geometry and a compact control structure for editing. A central challenge is deciding where to place anchors: raster appearance alone does not determine how a contour should be divided into B\'ezier segments. We present AnchorFlow, which learns anchor placement from designer-authored SVGs to reconstruct accurate curves with sparse controls. Our key idea is a sparse anchor field that jointly encodes contour geometry and reference segment junctions, including those along smooth contours. An anchor decoder predicts explicit anchor proposals from features learned under field supervision. These proposals guide boundary-constrained fitting and local refinement to recover cubic B\'ezier paths. On clean single-path inputs, AnchorFlow achieves 99.52\% mean IoU while using 56.6\% fewer anchors on average than AdaVec, with lower boundary error and closer agreement with source-SVG anchor layouts. Under boundary perturbations, it maintains high fidelity with limited anchor growth. Integrated into a component-based pipeline, the same path module also produces compact, faithful full-image reconstructions. On four local-editing tasks, our outputs require less median active time and fewer actions than AdaVec and LIVE while retaining high target-shape accuracy.
\end{abstract}
\input{sections/introduction}
\input{sections/method}
\input{sections/experiments}

\input{sections/conclusion_revised}

\FloatBarrier
\subsection*{AI use statement}
Generative AI tools assisted with experiment-design feedback, code implementation and review,  literature and format checks, translation, and manuscript drafting and editing. Numerical claims were checked against saved experimental outputs. The authors take responsibility for the manuscript and results.

\subsection*{Reproducibility statement}
Sections~\ref{app:ablation-details} and~\ref{app:implementation} of the Supplementary Material specify the architecture, training, reconstruction, evaluation, and ablation protocols. The accompanying code provides pretrained inference weights, single-path reconstruction and evaluation on 2,400 clean inputs, and six full-image examples. Its reproduction scope is detailed in the README.

\bibliography{refs}
\bibliographystyle{anchorflow}
\clearpage
\appendix
\input{sections/supplementary}
\end{document}

%% file: sections/introduction.tex
\section{Introduction}

Scalable Vector Graphics (SVG) preserve geometric structure, allowing artwork to be scaled and edited through its shapes and control points. Image vectorization reconstructs this representation from raster images, making existing artwork available for further editing. In icons and illustrations, closed paths often describe the outlines of individual components. Reconstructing such images involves organizing components and their layer order, as well as recovering the curve structure of each component. Even when component boundaries are available, recovering accurate paths with compact, editable controls remains a distinct challenge. We address this path-level reconstruction problem.

\begin{figure}[!ht]
\centering
\includegraphics[width=0.9\linewidth]{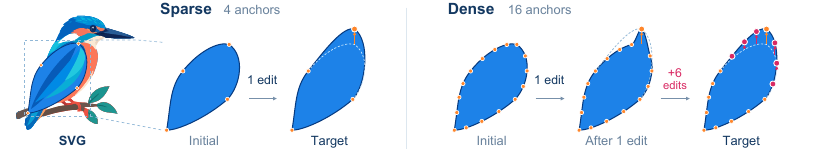}
\caption{\textbf{Anchor layout affects editing.}
The same initial wing outline is represented with 4 or 16 anchors. The illustrated adjustment involves one anchor group in the sparse representation and seven groups ($1+6$) in the dense representation.}
\label{fig:editing}
\end{figure}

Visually similar paths can nevertheless have very different control structures. Anchors, the junctions between consecutive curve segments, determine how these segments are distributed along an outline. Figure~\ref{fig:editing} illustrates how this affects editing. Both paths reproduce the same initial wing outline, but moving one anchor group reshapes a broad portion of the sparse contour, whereas the illustrated adjustment involves seven groups in the dense path. Coordinating more controls can make such adjustments laborious. Simply removing anchors, however, can lose corners, narrow features, or changes in curvature. How can we reconstruct a contour accurately while keeping its controls compact and manageable?

We approach this question through the segmentation of the observed contour. For a closed B\'ezier path, consecutive anchors specify the endpoints of each segment. Selecting and ordering an initial anchor set along the contour therefore determines both the number of segments and the portion of the contour assigned to each one. Once this segmentation is specified, each curve can be fitted to the corresponding portion of the input image boundary. Our intuition is to predict these segment junctions first, then reconstruct the curves between them using the observed boundary shape.

The same contour admits multiple segmentations, so visual agreement alone does not uniquely determine these junctions. Designer-authored SVGs provide examples of how shapes are divided into curve segments, including junctions on smooth arcs without visible corners. We introduce \method{} to learn a contour-conditioned anchor-layout prior from these examples. Its key representation is a \emph{sparse anchor field} that combines contour support with localized peaks at reference anchor positions. An anchor decoder reads features learned under field supervision to predict explicit anchor proposals. These proposals guide contour segmentation, while the observed boundary constrains curve fitting and refinement. This combination of learned segmentation and geometric fitting forms a reusable path reconstruction module, which we apply to the extracted components in our full-image pipeline.

We evaluate reconstruction quality, control structure, and editing performance. On 2,400 clean paths, \method achieves 99.52\% mean IoU with 9.53 anchors on average, using 56.6\% fewer anchors than AdaVec~\citep{zhao2025less} and agreeing more closely with source-SVG anchor layouts. It maintains high fidelity with limited anchor growth under boundary perturbations and produces compact full-image reconstructions. Our ablation studies further show that learning where to place anchors improves reconstruction accuracy while keeping the resulting paths compact. In an 18-participant study of four selected local-editing tasks, our outputs yield lower median active time and fewer actions than AdaVec and LIVE~\citep{ma2022live}, while retaining high target-shape accuracy. Our contributions are:

\begin{itemize}
\setlength{\itemsep}{2pt}
\setlength{\topsep}{3pt}
\setlength{\parsep}{0pt}
\setlength{\partopsep}{0pt}

\item \textbf{Learning anchor placement through a sparse field.}
We introduce a spatial representation that jointly encodes contour geometry and reference segment junctions, enabling anchor-layout learning from designer-authored SVGs, including junctions along smooth arcs.

\item \textbf{Anchor-guided SVG reconstruction.}
We combine learned anchor proposals with boundary-constrained curve fitting and bounded refinement to reconstruct accurate, compact B\'ezier paths, and extend this approach to full-image SVG reconstruction.

\item \textbf{Evaluation of reconstruction and editing.}
We demonstrate faithful, compact reconstruction on clean and perturbed paths and full images, supported by controlled placement comparisons and an editing study on four selected tasks.

\end{itemize}

\section{Related Work}
\label{sec:related_work}

Image vectorization reconstructs the shapes in a raster image as vector paths. Existing methods approach this task through boundary fitting and raster optimization, component decomposition and adaptive simplification, or learned vector representations.

\paragraph{Boundary fitting and raster optimization.}
Tracing methods such as Potrace, AutoTrace, and VTracer extract bitmap boundaries and convert them into vector paths through curve approximation and simplification~\citep{selinger2003potrace,autotrace,vtracer}. They construct curves using local boundary geometry and approximation error. Optimization-based methods render a candidate SVG and adjust its parameters to reduce disagreement with the input image. DiffVG provides differentiable rasterization for this process~\citep{li2020diffvg}; LIVE progressively adds and optimizes paths~\citep{ma2022live}, Optimize \& Reduce removes less important shapes during optimization~\citep{hirschorn2024optimize}, and Vector Scaffolding organizes optimization across scales~\citep{lee2026scaffolding}. These methods provide direct geometric or image-based constraints on shape reconstruction. Control structures depend on path initialization, curve budgets, and simplification criteria: similar reconstructed appearances do not imply the same curve segmentation or anchor layout.

\paragraph{Component decomposition and adaptive simplification.}
Full-image vectorization must also determine which regions should be represented as separate components or paths. SuperSVG groups neighboring pixels with similar appearance into superpixel regions and reconstructs vector graphics within them~\citep{hu2024supersvg}; SAMVG and SGLIVE use image segmentation to guide path construction~\citep{zhu2023samvg,zhou2024sglive}. These approaches provide component-level organization for complex images. AdaVec further combines component decomposition with adaptive control-point simplification: it traces component contours, merges neighboring curve segments using geometric criteria, and performs subsequent optimization~\citep{zhao2025less}. Adaptive vectorization therefore already addresses redundancy in both paths and control points. Our full-image pipeline uses an AdaVec-based component decomposition frontend and applies our path reconstruction module to each extracted component. Within each component, we investigate how segmentation regularities learned from designer-authored SVGs can provide an anchor-layout initialization for subsequent geometric fitting and refinement.

\paragraph{Learned vector representations and reconstruction.}
Learning from vector examples provides data-driven priors for path organization and control-point placement. SVG-VAE and DeepSVG learn structural regularities by modeling drawing commands and their organization~\citep{lopes2019learned,carlier2020deepsvg}; StarVector, OmniSVG, and InternSVG further use multimodal models to generate SVG code or command sequences from input images~\citep{rodriguez2025starvector,yang2025omnisvg,wang2026internsvg}. These advances show that designer-authored SVGs can provide structural priors beyond local boundary geometry. However, direct SVG generation predicts both path structure and curve coordinates, and command-sequence prediction does not itself explicitly constrain individual curves to follow the observed boundary. Consequently, a result with coherent structure and similar overall appearance can still exhibit local contour shifts or distorted details. \method{} learns an anchor-layout prior from reference SVGs and combines it with boundary-constrained curve fitting and refinement, allowing learned segmentation patterns to guide the control structure while the observed contour constrains the geometry.

\begin{figure}[!t]
\centering
\includegraphics[width=0.92\linewidth]{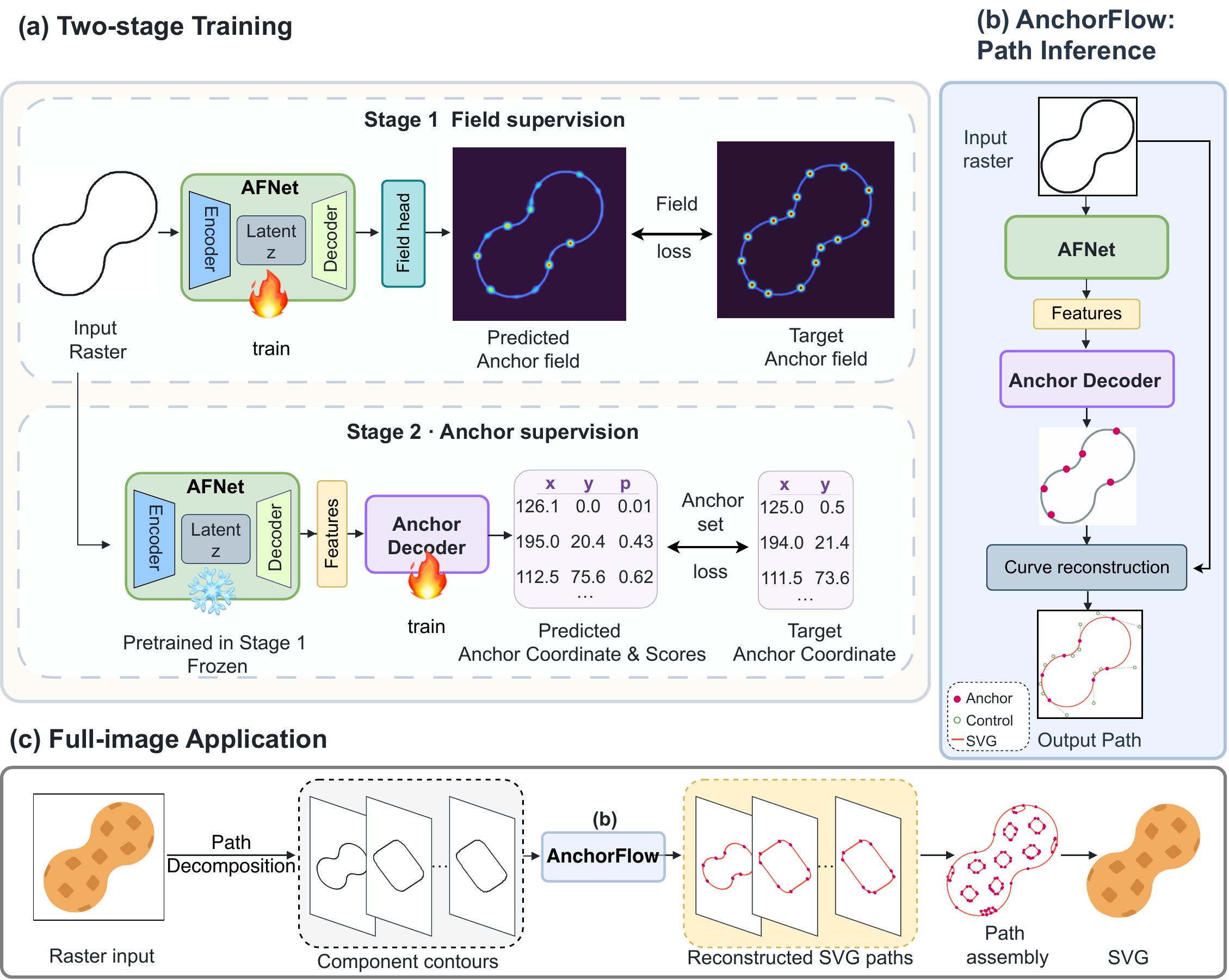}
\caption{\textbf{Overview.}
(a) Field supervision trains AFNet to encode contour geometry and reference anchor locations. The anchor decoder is then trained on frozen AFNet features to predict coordinates and existence scores.
(b) At inference, the predicted scores guide anchor selection, while the observed contour determines anchor order and provides boundary samples for curve fitting.
(c) Full-image reconstruction: reconstruct component paths and assemble them into an SVG.}
\label{fig:pipeline}
\end{figure}

%% file: sections/method.tex
\section{Method}
\label{sec:method}

\method{} reconstructs a compact, closed cubic B\'ezier path from a rasterized component contour. Field supervision learns contour and anchor-layout features, an anchor decoder predicts segment junctions, and boundary-constrained fitting and refinement recover the curves between them. Figure~\ref{fig:pipeline} summarizes the pipeline.

\subsection{Anchor-field learning}
\label{sec:anchor_field}

\paragraph{Representing shape and segment junctions.}
We rasterize the observed boundary into a $512\times512$ contour image $X$ as input to AFNet. From the corresponding clean source contour $\Gamma$ and its source-SVG anchor set $A^*$, we construct a target field $F^*$ that combines a low-amplitude contour band $c_\Gamma$ with Gaussian anchor peaks:
\begin{equation}
F^*(u)=
\max\left\{
c_\Gamma(u),
\max_{a\in A^*}
\exp\left(-\frac{\|u-a\|_2^2}{2\sigma_a^2}\right)
\right\}.
\label{eq:field}
\end{equation}
Here $u$ is a pixel coordinate, $a$ is a reference anchor, and $\sigma_a$ controls the anchor-peak width. The contour band supplies boundary context, while the peaks mark the reference segment junctions.

\paragraph{Field supervision.}
AFNet is a convolutional encoder--decoder with skip connections that preserve spatial detail. With network parameters $\theta$, it predicts $F_\theta(X)$ and is trained using
\begin{equation}
\mathcal L_{\rm field}
=
\frac{1}{HW}\sum_u
\left(5e_u^2+|e_u|\right),
\qquad
e_u=F_\theta(X)_u-F^*(u),
\label{eq:field_loss}
\end{equation}
where $H$ and $W$ are the field height and width, and $e_u$ is the prediction error at pixel $u$. For perturbed inputs, target fields are constructed from the corresponding clean source contours and anchors using the same geometric transform as the input.

\subsection{Anchor-set decoding}
\label{sec:anchor_decoding}

\paragraph{Reading anchor proposals from features.}
The predicted anchor field supports direct peak extraction. However, the number and locations of extracted anchors depend on response thresholds and the nonmaximum-suppression and centroid windows. We therefore use a three-layer attention-based decoder to predict anchor coordinates and existence scores from frozen AFNet features, providing a learned criterion for candidate selection. Intermediate features supply spatial detail, while pooled and bottleneck features supply global context. The field-output head is unused during decoder training and inference. We use this decoder throughout reconstruction and evaluate alternative readouts in Section~\ref{sec:ablation}.

The decoder maintains learned queries, each paired with a candidate coordinate. Each layer samples fine features around these coordinates, exchanges information among queries through self-attention, and reads global context through cross-attention. Prediction heads then update the normalized coordinate $r_j\in[0,1]^2$ and existence score $p_j\in[0,1]$ for query $j$. Supplementary Figure~\ref{fig:network_details} details the feature connections and decoder architecture.

\paragraph{Supervising anchor presence and position.}
A closed contour has no preferred starting anchor or traversal direction. We therefore supervise candidate anchors as a set, without assigning query indices to a prescribed contour order. Let $\bar a_i^*$ denote reference anchor $a_i^*$ normalized to $[0,1]^2$. Bipartite matching uses coordinate $L_1$ distance and focal classification cost to assign each reference anchor to a query $\pi(i)$. Matched queries receive existence label $y_j=1$, and unmatched queries receive $y_j=0$. At each layer, we optimize
\begin{equation}
\mathcal L_{\rm set}
=
\mathcal L_{\rm focal}(p,y)
+
25\,\mathcal L_{\rm SmoothL1}
\left(\{r_{\pi(i)}\},\{\bar a_i^*\}\right),
\label{eq:setloss}
\end{equation}
where $p=(p_j)$ and $y=(y_j)$. The focal loss supervises existence for all queries, while the Smooth-$L_1$ loss supervises coordinates for matched queries. Losses are normalized by the target count and averaged across layers. At inference, candidates with final scores $p_j\geq\tau$ are retained, where $\tau$ is the selection threshold, allowing the initial anchor count to vary across inputs.

\subsection{Boundary-constrained curve reconstruction}
\label{sec:curve_reconstruction}

\paragraph{From anchors to fitted curves.}
We assign selected proposals to distinct locations on the observed contour and order them by boundary arclength. Consecutive anchors define the endpoints and boundary span of each curve. For segment $i$, we fix endpoints $a_i$ and $a_{i+1}$ and fit a cubic B\'ezier curve $B_i(t;\mathbf H_i)$, where $\mathbf H_i$ collects its two inner control points. The curve parameterization is given in Supplementary Eq.~\eqref{eq:bezier}.

We fit each segment to boundary samples $x_{im}$ at fixed curve parameters $t_{im}$, using coordinates centered at the midpoint of its endpoints:
\begin{equation}
\widehat{\mathbf H}_i
=
\arg\min_{\mathbf H_i}
\sum_m
\left\|B_i(t_{im};\mathbf H_i)-x_{im}\right\|_2^2
+
\lambda_i\|\mathbf H_i\|_F^2,
\label{eq:handle_fit}
\end{equation}
where $\lambda_i>0$ is the regularization weight and $\|\cdot\|_F$ denotes the Frobenius norm.

For single-path reconstruction, failed geometric checks or insufficient IoU with the observed input lead us to lower the selection threshold, including more anchor proposals in curve reconstruction.

\paragraph{Removing redundant anchors.}
After obtaining a valid path, we check for redundant intermediate anchors. For consecutive anchors $a,b,c$, we refit one curve from $a$ to $c$ against their combined boundary span. We remove $b$ only if the replacement passes sampled self-intersection checks and satisfies fixed tolerances on raster fidelity and geometric deviation.

For full-image reconstruction, an AdaVec-based frontend extracts component contours~\citep{zhao2025less}. We apply the path reconstruction procedure to each component, map the resulting paths back to the image canvas, and assemble them with their fills and layer order before global optimization.

%% file: sections/experiments.tex
\section{Experiments}

\subsection{Experimental setup}
\label{sec:setup}

\paragraph{Data and training.}
We train \method on 31,950 paths extracted from 15,996 SVG files. Sources include SVG data from StarVector~\citep{rodriguez2025starvector}, Fluent Emoji~\citep{fluentemoji}, Noto Emoji~\citep{notoemoji}, Game-icons, Health Icons, Material Design Icons, Font Awesome, and Natural Earth. Training uses clean and perturbed contour views with anchor annotations from the source SVGs. Both training stages use a single NVIDIA A100 80GB GPU, taking approximately 9.6 hours for AFNet and 5.0 hours for the anchor decoder (14.6 hours total).

Single-path evaluation uses 2,400 paths from ColorSVG~\citep{chen2025svgbuilder}, Bootstrap Icons~\citep{bootstrapicons}, and STIX~\citep{stixfonts}. For each path, we evaluate a clean raster and one fixed boundary-perturbed input against the same clean reference.
Full-image evaluation randomly samples 100 images each from Noto Emoji, Fluent Emoji, and HeisenVec~\citep{zini2025vhector}. Source SVG files used to extract training paths are excluded from both single-path and full-image evaluation.

\paragraph{Baselines.}
We compare tracing methods Potrace~\citep{selinger2003potrace}, VTracer~\citep{vtracer}, and AutoTrace~\citep{autotrace}; optimization-based methods DiffVG~\citep{li2020diffvg} and LIVE~\citep{ma2022live}; SVG generators OmniSVG~\citep{yang2025omnisvg} and InternSVG~\citep{wang2026internsvg}; and AdaVec~\citep{zhao2025less} for adaptive parameterization.

\paragraph{Metrics.}
For single paths, we report region IoU, symmetric boundary Chamfer distance (CD, in pixels), and anchor count.
Count MAE measures mean absolute anchor-count error relative to the source SVG, while GOSPA~\citep{rahmathullah2017gospa} measures the discrepancy between reconstructed and source-SVG anchor sets. 
For full images, RGB MSE and PSNR measure pixel-level fidelity, LPIPS~\citep{zhang2018lpips} measures perceptual differences, and SSIM~\citep{wang2004ssim} measures structural similarity. Comp.\ denotes the percentage of completed outputs. 
\input{tables/table12_clean_perturbed} 
\begin{figure}[!htbp]
\centering
\includegraphics[width=0.9\linewidth]{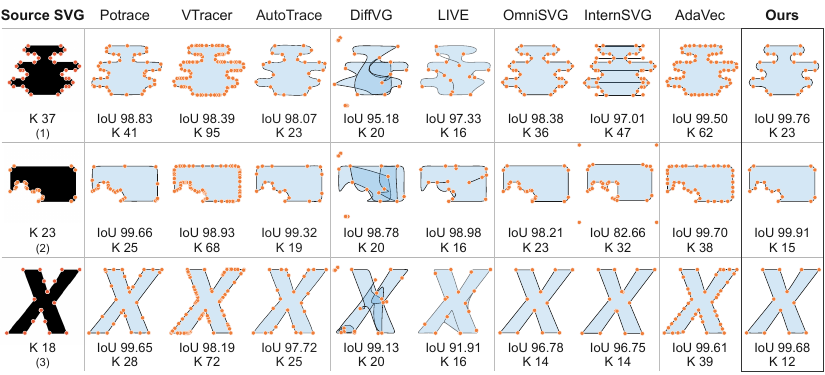}
\caption{\textbf{Clean single-path reconstruction.}
The first column shows black input raster shapes with orange source-SVG anchors.
IoU (\%) and anchor count $K$ appear below each reconstruction.}
\label{fig:single}
\end{figure}
\subsection{Single-path reconstruction}
\label{sec:singlepath}
\paragraph{Clean reconstruction.}
On clean inputs, \method achieves the highest mean IoU (99.521\%) and the lowest CD (0.072 pixels), anchor count (9.53), count MAE, and GOSPA (Table~\ref{tab:table12_clean_perturbed}). It improves fidelity with 56.6\% fewer anchors than AdaVec; even the most compact tracer, AutoTrace, uses 12.97 anchors.

Figure~\ref{fig:single} illustrates the reconstructed control structures. Example~(1) preserves indentations and straight-to-curve transitions with sparser anchors than VTracer and AdaVec. Example~(2) captures the curved profile at 99.91\% IoU with 15 anchors, versus 38 for AdaVec. In example~(3), 12 anchors capture the X-shaped contour's straight edges and sharp corners at 99.68\% IoU, versus 28 for Potrace and 39 for AdaVec. Together, these results show that compact control structures can preserve both smooth contours and sharp local features.

\paragraph{Robustness to boundary perturbations.}
Segmentation, resizing, and rasterization can introduce boundary irregularities. We therefore test robustness using lower-resolution rasterization and weak boundary displacement. In the perturbed results of Table~\ref{tab:table12_clean_perturbed}, \method maintains 97.687\% mean IoU and a CD of 0.317 pixels with 12.55 anchors on average. Relative to clean inputs, its IoU decreases by 1.834 percentage points, versus 4.262 for AdaVec, while its anchor count grows by 31.7\%, versus 68.1\%. Potrace achieves a comparable IoU of 97.523\%, but its anchor count grows by 238.4\%. LIVE falls from 94.930\% to 63.575\% mean IoU, indicating sensitivity to these perturbations. Thus, the fidelity--compactness advantage persists under these perturbations: \method preserves the underlying shape without the large anchor growth observed in the tracing baselines.

\begin{figure}[!t]
\centering
\includegraphics[width=0.95\linewidth]{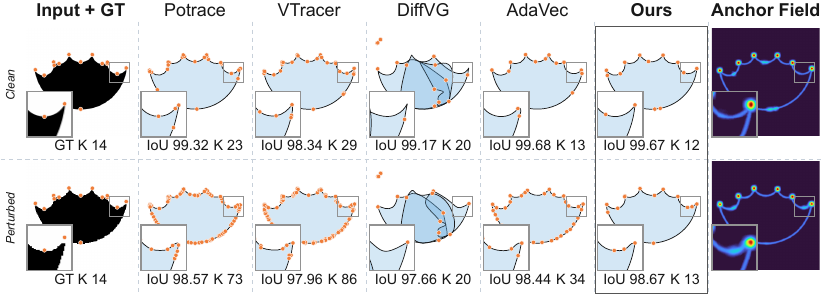}
\caption{\textbf{Reconstruction under boundary perturbations.}
Clean and perturbed inputs and reconstructions appear in the top and bottom rows, respectively. AFNet fields on the right.}
\label{fig:perturbation}
\end{figure}

\begin{figure}[!t]
\centering
\includegraphics[width=0.9\linewidth]{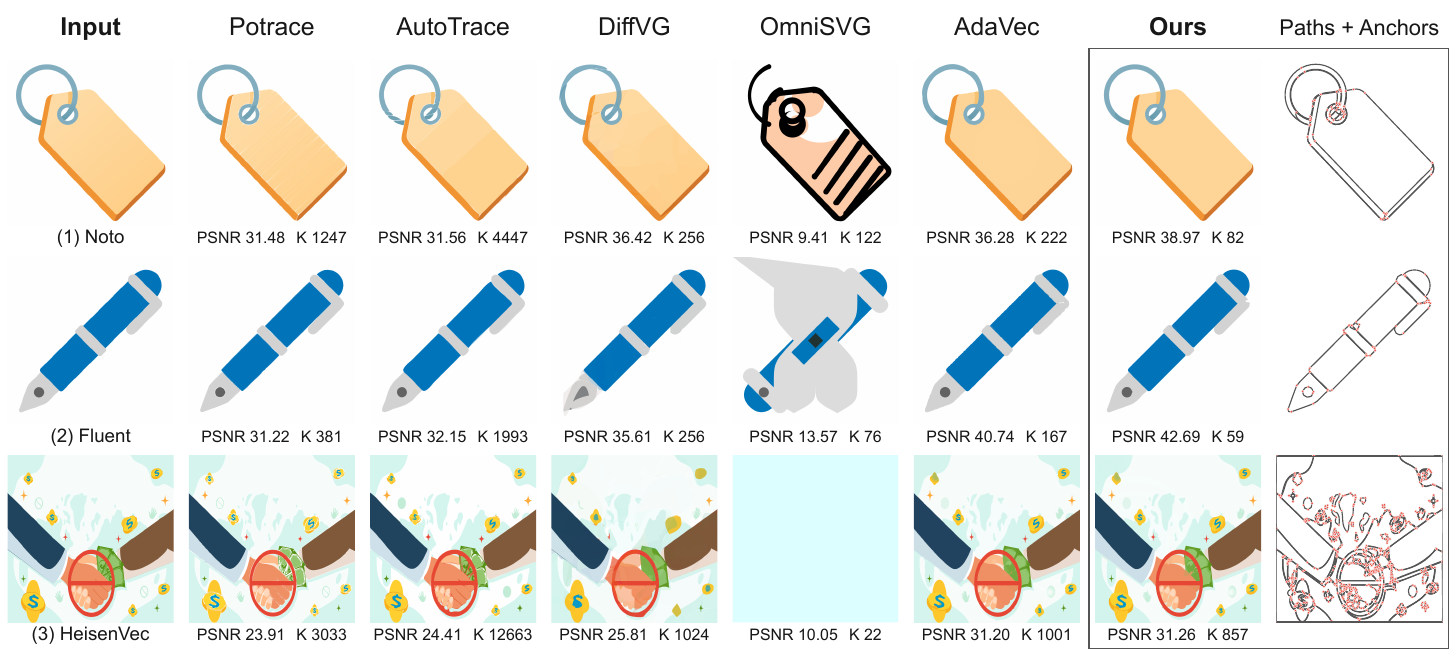}
\caption{\textbf{Full-image reconstruction.}
Qualitative comparisons on Noto Emoji, Fluent Emoji, and HeisenVec. 
The last column visualizes our SVG paths and anchors.}
\label{fig:full}
\end{figure}

\subsection{Full-image reconstruction}
\input{tables/table4_fullimage300}
Applied to full-image components, \method uses the fewest anchors on all three datasets and leads on the four fidelity metrics, except for second-best MSE on Fluent Emoji and LPIPS on HeisenVec (Table~\ref{tab:table4_fullimage300}). Anchor counts on Noto and Fluent are 31.1\% and 55.4\% below the next most compact baseline. Figure~\ref{fig:full} shows higher PSNR with substantially fewer anchors than Potrace and AutoTrace; OmniSVG exhibits larger shape and layout deviations. In example~(2), \method preserves the pointed outline and circular feature where DiffVG distorts the pen tip. Our outputs remain visually close to AdaVec's with fewer anchors, including the complex scene in example~(3). These results show that the compactness and fidelity of the path module carry over to assembled images with multiple components.

\subsection{Ablation study}
\label{sec:ablation}
We examine what the learned representation and anchor placement contribute. Table~\ref{tab:readout-ablation} compares readout and training choices; Tables~\ref{tab:shared-backend} and~\ref{tab:matched-anchor-placement} control the reconstruction backend and initial count to examine the value of anchor initialization and placement.
\input{tables/ablx2}
\input{tables/matched_anchor_placement}

\paragraph{Anchor readout and field pretraining.}
We compare field-peak extraction and the anchor decoder using the same field-pretrained AFNet. In Table~\ref{tab:readout-ablation}, field peaks achieve slightly higher IoU, while the decoder uses 3.4\% fewer final anchors and has 0.9\% lower GOSPA. The field features therefore support accurate reconstruction through either readout, with the decoder favoring a more compact anchor structure.

Under the same total training budget of 120 epochs, we compare our two-stage scheme (60 field-training and 60 decoder-training epochs) with joint AFNet--decoder training from scratch using anchor-set supervision alone. Our scheme yields fewer anchors, slightly higher IoU and lower GOSPA, with shorter recorded training time (14.6 vs. 26.5 hours). Field pretraining thus provides a useful initialization for compact reconstruction within the same epoch budget.

\paragraph{Effect of anchor initialization.}
We compare initial anchors from Ours, Potrace, and AdaVec under a shared geometric backend. We discard the handles produced by Potrace and AdaVec and apply the same curve reconstruction to the observed contours; Ours retains its standard adaptive anchor selection. AdaVec endpoints are taken after simplification but before point and shape optimization. Table~\ref{tab:shared-backend} shows that Ours yields the fewest final anchors and closer agreement with source-SVG anchors while maintaining comparable fidelity. These results show that the initial anchors affect the final control structure even after the same geometric refinement.

Initial anchor count also affects reconstruction accuracy. To isolate the effect of anchor position, we compare Ours with uniform arclength sampling (Uniform) and greedy insertion by fitting error (Error-greedy), matching their initial counts to Ours for each input. With anchors fixed and only handles fitted, Ours already achieves higher IoU and lower GOSPA (Table~\ref{tab:matched-anchor-placement}, A), showing that its anchor positions enable more accurate curves with the same number of segments. This advantage persists after shared refinement (B): compared with Error-greedy, Ours improves mean IoU by 0.40 percentage points, with 3.6\% fewer final anchors and 36.5\% lower GOSPA. These results show that the anchor positions predicted by Ours make more effective use of the initial anchor budget and lead to more accurate, compact curves after refinement.

\subsection{Editing use-case study}
\label{sec:usecase}
\begin{figure}[!htbp]
\centering
\includegraphics[width=0.9\linewidth]{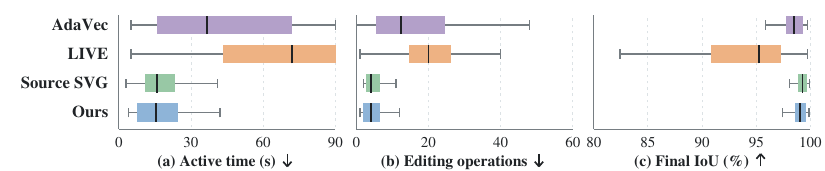}
\caption{\textbf{User study on four SVG editing tasks.} Eighteen participants perform four editing tasks under four conditions: AdaVec, \method, LIVE, and the source SVG as a reference.}
\label{fig:usecase}
\end{figure}

To examine whether the reconstructed controls support practical editing, we evaluate four distinct editing tasks with 18 participants under four conditions: native AdaVec~\citep{zhao2025less}, \method, LIVE~\citep{ma2022live}, and the source SVG as a reference. Across the four tasks, \method achieves editing performance close to the source-SVG reference, requiring a similar median active editing time (15.5 vs. 16.0 seconds) and the same median number of geometric actions (4), while reaching a comparable final mask IoU (99.04\% vs. 99.27\%). Compared with AdaVec and LIVE, \method requires less time and fewer actions while achieving higher final target IoU on every task. The compact reconstructed controls therefore translate into reduced editing effort while preserving target accuracy on these tasks.

%% file: tables/table12_clean_perturbed.tex
\begin{table}[!t]
\centering
\tablecaptiongap
\caption{Clean and perturbed reconstruction of 2,400 paths. Best results are bold; second-best are underlined. $N$ is the path budget,
$T$ is the maximum output-token count. }
\label{tab:table12_clean_perturbed}
\begingroup
\setlength{\tabcolsep}{2pt}
\setlength{\aboverulesep}{1pt}
\setlength{\belowrulesep}{1pt}
\renewcommand{\arraystretch}{1.0}
\footnotesize
\begin{tabular}{@{}cllrrrrrrr@{}}
\toprule
\textbf{Input} & \textbf{Type} & \textbf{Method} & \shortstack{\textbf{Comp.}\\\textbf{(\%) $\uparrow$}} & \shortstack{\textbf{IoU}\\\textbf{(\%) $\uparrow$}} & \shortstack{\textbf{CD}\\\textbf{(px) $\downarrow$}} & \textbf{Anchors}$\downarrow$ & \shortstack{\textbf{Count}\\\textbf{MAE $\downarrow$}} & \textbf{GOSPA}$\downarrow$ & \shortstack{\textbf{Time}\\\textbf{(s) $\downarrow$}} \\
\midrule
\multirow{9}{*}[-3pt]{\rotatebox[origin=c]{90}{Clean}} & \multirow{3}{*}{Tracing} & Potrace & 100.0 & 99.168 & 0.109 & 17.88 & 6.745 & 0.0562 & \underline{0.0146} \\
 &  & VTracer & 100.0 & 97.664 & 0.314 & 36.83 & 25.243 & 0.0774 & \textbf{0.0036} \\
 &  & AutoTrace & 100.0 & 98.073 & 0.257 & \underline{12.97} & \underline{3.565} & \underline{0.0501} & 0.0711 \\
\cmidrule[.25pt](lr){2-10}
 & \multirow{2}{*}{Opt.} & DiffVG {\scriptsize ($N=5$)} & 100.0 & 89.424 & 18.738 & 20.00 & 10.946 & 0.0725 & 24.52 \\
 &  & LIVE {\scriptsize ($N=1$)} & 100.0 & 94.930 & 4.132 & 16.00 & 8.305 & 0.0653 & 52.31 \\
\cmidrule[.25pt](lr){2-10}
 & \multirow{2}{*}{Gen.} & OmniSVG {\scriptsize ($T=1{,}024$)} & 86.2 & 59.657 & 44.593 & 22.63 & 13.194 & 0.0475 & 5.47 \\
 &  & InternSVG {\scriptsize ($T=1{,}024$)} & 94.1 & 64.897 & 24.993 & 14.53 & 6.642 & 0.0498 & 4.77 \\
\cmidrule[.25pt](lr){2-10}
 & \multirow{2}{*}{Adapt.} & AdaVec & 100.0 & \underline{99.254} & \underline{0.087} & 21.95 & 11.340 & 0.0620 & 9.65 \\
 &  & \textbf{Ours} & 100.0 & \textbf{99.521} & \textbf{0.072} & \textbf{9.53} & \textbf{2.663} & \textbf{0.0254} & 0.836 \\
\midrule
\multirow{9}{*}[-3pt]{\rotatebox[origin=c]{90}{Perturbed}} & \multirow{3}{*}{Tracing} & Potrace & 100.0 & \underline{97.523} & \underline{0.334} & 60.51 & 48.758 & 0.1023 & \underline{0.266} \\
 &  & VTracer & 100.0 & 96.456 & 0.495 & 96.63 & 84.837 & 0.1307 & \textbf{0.252} \\
 &  & AutoTrace & 100.0 & 96.596 & 0.468 & 18.83 & \underline{8.228} & \underline{0.0607} & 0.475 \\
\cmidrule[.25pt](lr){2-10}
 & \multirow{2}{*}{Opt.} & DiffVG {\scriptsize ($N=5$)} & 100.0 & 87.976 & 19.397 & 20.00 & 10.946 & 0.0734 & 53.65 \\
 &  & LIVE {\scriptsize ($N=1$)} & 100.0 & 63.575 & 83.903 & \underline{16.00} & 8.305 & 0.0688 & 83.13 \\
\cmidrule[.25pt](lr){2-10}
 & \multirow{2}{*}{Gen.} & OmniSVG {\scriptsize ($T=1{,}024$)} & 14.5 & 8.363 & 222.394 & 159.93 & 32.226 & 0.0645 & 45.76 \\
 &  & InternSVG {\scriptsize ($T=1{,}024$)} & 20.6 & 6.886 & 217.258 & 52.70 & 19.317 & 0.0569 & 42.56 \\
\cmidrule[.25pt](lr){2-10}
 & \multirow{2}{*}{Adapt.} & AdaVec & 97.3 & 94.992 & 7.144 & 36.89 & 25.195 & 0.0813 & 11.07 \\
 &  & \textbf{Ours} & 100.0 & \textbf{97.687} & \textbf{0.317} & \textbf{12.55} & \textbf{3.470} & \textbf{0.0376} & 2.13 \\
\bottomrule
\end{tabular}
\endgroup
\end{table}

%% file: tables/table4_fullimage300.tex
\begin{table}[!htbp]
\centering
\tablecaptiongap
\caption{\textbf{Full-image reconstruction.}
Quantitative comparison of representation compactness, reconstruction quality, and runtime on Noto Emoji, Fluent Emoji, and HeisenVec (100 images each). }
\label{tab:table4_fullimage300}
\begingroup
\setlength{\tabcolsep}{3pt}
\setlength{\aboverulesep}{1pt}
\setlength{\belowrulesep}{1pt}
\renewcommand{\arraystretch}{1.0}
\footnotesize
\begin{tabular}{@{}lllrrrrrr@{}}
\toprule
\textbf{Dataset} & \textbf{Type} & \textbf{Method} & \textbf{Anchors}$\downarrow$ & \textbf{MSE}$\downarrow$ & \textbf{LPIPS}$\downarrow$ & \textbf{PSNR}$\uparrow$ & \textbf{SSIM}$\uparrow$ & \textbf{Time (s)} \\
\midrule
\multirow{6}{*}[-3pt]{\shortstack[l]{Noto\\Emoji}} & \multirow{2}{*}{Tracing} & Potrace (color) & 1732.45 & 0.003768 & 0.09187 & 25.34 & 0.9131 & \textbf{0.22} \\
 &  & AutoTrace & 6262.27 & 0.008919 & 0.06353 & 26.95 & 0.9479 & \underline{0.33} \\
\cmidrule[.25pt](lr){2-9}
 & Opt. & DiffVG {\scriptsize ($N=64$)} & \underline{256.00} & 0.001911 & 0.07013 & 30.00 & 0.9550 & 41.47 \\
\cmidrule[.25pt](lr){2-9}
 & Gen. & OmniSVG {\scriptsize ($T=4{,}096$)} & 603.93 & 0.139472 & 0.49974 & 10.20 & 0.7179 & 69.59 \\
\cmidrule[.25pt](lr){2-9}
 & \multirow{2}{*}{Adapt.} & AdaVec & 281.13 & \underline{0.001287} & \underline{0.02918} & \underline{34.66} & \underline{0.9787} & 31.17 \\
 &  & \textbf{Ours} & \textbf{176.49} & \textbf{0.000484} & \textbf{0.01298} & \textbf{37.32} & \textbf{0.9872} & 37.09 \\
\midrule
\multirow{6}{*}[-3pt]{\shortstack[l]{Fluent\\Emoji}} & \multirow{2}{*}{Tracing} & Potrace (color) & 318.98 & 0.001478 & 0.03161 & 29.00 & 0.9718 & \textbf{0.19} \\
 &  & AutoTrace & 4167.44 & 0.010476 & 0.07829 & 26.83 & 0.9635 & \underline{0.28} \\
\cmidrule[.25pt](lr){2-9}
 & Opt. & DiffVG {\scriptsize ($N=64$)} & 256.00 & 0.000718 & 0.02757 & 36.12 & 0.9856 & 40.88 \\
\cmidrule[.25pt](lr){2-9}
 & Gen. & OmniSVG {\scriptsize ($T=4{,}096$)} & 298.08 & 0.071542 & 0.29301 & 15.10 & 0.8349 & 33.11 \\
\cmidrule[.25pt](lr){2-9}
 & \multirow{2}{*}{Adapt.} & AdaVec & \underline{157.49} & \textbf{0.000180} & \underline{0.00930} & \underline{39.94} & \underline{0.9949} & 24.89 \\
 &  & \textbf{Ours} & \textbf{70.27} & \underline{0.000195} & \textbf{0.00558} & \textbf{41.89} & \textbf{0.9956} & 41.21 \\
\midrule
\multirow{6}{*}[-3pt]{HeisenVec} & \multirow{2}{*}{Tracing} & Potrace (color) & 3314.02 & 0.004243 & \textbf{0.06673} & 24.51 & 0.9180 & \textbf{0.37} \\
 &  & AutoTrace & 10992.80 & 0.006089 & 0.07671 & 23.65 & 0.9179 & \underline{0.38} \\
\cmidrule[.25pt](lr){2-9}
 & Opt. & DiffVG {\scriptsize ($N=256$)} & 1024.00 & 0.002824 & 0.10758 & 26.14 & 0.9163 & 148.01 \\
\cmidrule[.25pt](lr){2-9}
 & Gen. & OmniSVG {\scriptsize ($T=8{,}192$)} & 1312.51 & 0.126347 & 0.60392 & 10.85 & 0.6887 & 186.25 \\
\cmidrule[.25pt](lr){2-9}
 & \multirow{2}{*}{Adapt.} & AdaVec & \underline{926.84} & \underline{0.002152} & 0.07899 & \underline{27.82} & \underline{0.9408} & 79.96 \\
 &  & \textbf{Ours} & \textbf{892.11} & \textbf{0.001779} & \underline{0.07206} & \textbf{28.25} & \textbf{0.9449} & 172.64 \\
\bottomrule
\end{tabular}
\endgroup
\end{table}

%% file: tables/ablx2.tex
\begin{table}[t]
\centering
\begin{minipage}[t]{0.49\linewidth}
\vspace{0pt}
\centering
\caption{\textbf{Readout and field pretraining.}\newline
Means over 2,400 clean paths; all outputs valid.}
\label{tab:readout-ablation}
\fontsize{7.7}{9.5}\selectfont
\setlength{\tabcolsep}{0pt}
\renewcommand{\arraystretch}{1.08}
\begin{tabular*}{\linewidth}{@{\extracolsep{\fill}}lrrr@{}}
\toprule
Method
& $K\downarrow$
& IoU$\uparrow$
& GOSPA$\downarrow$ \\
\midrule
Field-peak readout
& 9.872 & \textbf{0.995679} & 0.025616 \\
Joint from scratch
& 9.670 & 0.995151 & 0.025869 \\
\textbf{Ours}
& \textbf{9.533}
& 0.995211
& \textbf{0.025378} \\
\bottomrule
\end{tabular*}
\end{minipage}\hfill
\begin{minipage}[t]{0.50\linewidth}
\vspace{0pt}
\centering
\caption{\textbf{Effect of anchor initialization.}\newline
Results on 2,400 clean paths.}
\label{tab:shared-backend}
\fontsize{7.7}{9.5}\selectfont
\setlength{\tabcolsep}{0pt}
\renewcommand{\arraystretch}{1.08}
\begin{tabular*}{\linewidth}{@{\extracolsep{\fill}}lrrrr@{}}
\toprule
Method
& Initial $K$
& Final $K\downarrow$
& IoU$\uparrow$
& GOSPA$\downarrow$ \\
\midrule
Potrace
& 17.877 & 12.370 & 0.994684 & 0.046554 \\
AdaVec
& 21.945 & 11.516 & 0.994856 & 0.046738 \\
\textbf{Ours}
& \textbf{11.585}
& \textbf{9.533}
& \textbf{0.995211}
& \textbf{0.025378} \\
\bottomrule
\end{tabular*}
\end{minipage}
\end{table}

%% file: tables/matched_anchor_placement.tex
\begin{table}[!htbp]
\centering
\small
\caption{\textbf{Anchor placement at matched counts.}
Results on 2,400 clean paths. A: handle fitting with fixed anchors. B: shared refinement, independently restarted from the same initializations.}
\label{tab:matched-anchor-placement}
\setlength{\tabcolsep}{0pt}
\renewcommand{\arraystretch}{1.04}
\begin{tabular*}{\linewidth}{@{\extracolsep{\fill}}lrrrrrr@{}}
\toprule
& & \multicolumn{2}{c}{A: Handle fitting only}
& \multicolumn{3}{c}{B: Shared refinement} \\
\cmidrule(lr){3-4}\cmidrule(l){5-7}
Method & Initial $K$ & IoU $\uparrow$ & GOSPA $\downarrow$
& Final $K$ $\downarrow$ & IoU $\uparrow$ & GOSPA $\downarrow$ \\
\midrule
Uniform
& 13.879
& 0.977403
& 0.058960
& 11.293
& 0.988979
& 0.052135 \\

Error-greedy
& 13.879
& 0.985210
& 0.047443
& 10.478
& 0.991846
& 0.040808 \\

\textbf{Ours}
& 13.879
& \textbf{0.993993}
& \textbf{0.030902}
& \textbf{10.098}
& \textbf{0.995833}
& \textbf{0.025928} \\
\bottomrule
\end{tabular*}
\end{table}

%% file: sections/conclusion_revised.tex
\section{Conclusion and limitations}
\label{sec:conclusion}
\label{sec:failures}

We presented \method, which learns anchor placement from designer-authored SVGs through a sparse field encoding contour geometry and reference anchors. An anchor decoder predicts explicit anchors for boundary-constrained curve fitting. Experiments show faithful, compact reconstruction on single paths and full images, alongside efficient editing on the selected tasks. The method remains sensitive to errors in component decomposition and anchor placement, which can omit components or distort fine details, and is slower than the evaluated tracing methods.

%% file: sections/supplementary.tex
\section*{Supplementary Material}
\section{Supplementary experiments and analyses}
\label{app:additional-experiments}

\subsection{Failure case study}
\label{app:failure-cases}

Figure~\ref{fig:failures} illustrates two failure modes. In (a), errors in the component decomposition omit or merge foreground regions, leaving details missing from the assembled image. In (b), the reconstructed rake preserves the overall silhouette, but the widths and local curvature of individual teeth deviate from the reference. The first case concerns the decomposition frontend; the second shows that local anchor placement and curve fitting can still distort fine geometry.

\begin{center}
\begin{minipage}{\linewidth}
\centering
\includegraphics[width=0.80\linewidth]{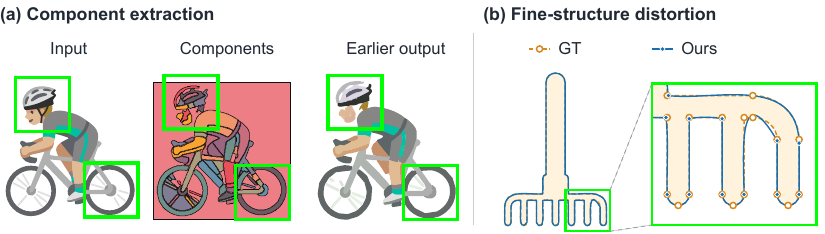}
\makeatletter\def\@captype{figure}\makeatother
\caption{\textbf{Failure cases.} (a) Component decomposition errors affecting full-image reconstruction. (b) Fine-detail distortion in single-path reconstruction.}
\label{fig:failures}
\end{minipage}
\end{center}

\subsection{Comparison with general-purpose models}
\label{app:llm}

We randomly sample 100 clean paths from the 2,400-path benchmark and compare GPT-6 Astra~\citep{openai2026gpt6astra}, GPT-5.6 Sol~\citep{openai2026gpt56}, Claude Fable 5.1~\citep{anthropic2026fable51}, Claude Opus 5~\citep{anthropic2026opus5}, DeepSeek-V4.1-Flash~\citep{deepseekai2026deepseekv41flash}, and Gemini 3.8 Flash~\citep{google2026gemini38flash}. All models receive the same inputs and produce one candidate per input.

\method achieves 99.617\% mean IoU with 9.74 anchors, compared with 97.778\% and 97.239\% IoU for Astra and Sol. It also has lower CD and GOSPA than all six baselines (Figure~\ref{fig:llm_comparison}). Model settings and measurable-output counts are given in Section~\ref{app:protocol}.

\begin{center}
\begin{minipage}{0.90\linewidth}
\centering
\begin{minipage}{\linewidth}
\begingroup
\def\AFsuppLLMLegend#1#2{%
  \includegraphics[width=.41\linewidth,
    trim=406.328740bp #1bp 5.846457bp #2bp,clip]
    {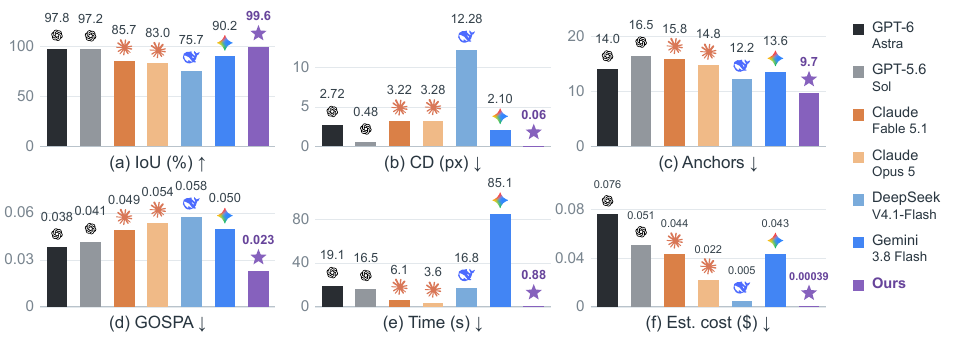}%
}
\includegraphics[width=\linewidth,
  trim=0bp 78.927165bp 66.649606bp 0bp,clip]
  {fig/llm_comparison_bottom_labels_20260917.pdf}\par
\noindent
\begin{minipage}[c]{.661808\linewidth}
  \includegraphics[width=\linewidth,
    trim=0bp 0bp 202.287401bp 84.188976bp,clip]
    {fig/llm_comparison_bottom_labels_20260917.pdf}
\end{minipage}%
\begin{minipage}[c]{.338192\linewidth}
  \centering
  \begin{tabular}{@{}c@{\hspace{.035\linewidth}}c@{}}
    \AFsuppLLMLegend{136.222441}{8.769685} &
    \AFsuppLLMLegend{115.759842}{29.232283} \\
    \AFsuppLLMLegend{95.297244}{49.694882} &
    \AFsuppLLMLegend{74.834646}{70.157480} \\
    \AFsuppLLMLegend{54.372047}{90.620079} &
    \AFsuppLLMLegend{33.909449}{111.082677} \\
    \AFsuppLLMLegend{13.446850}{131.545275} &
  \end{tabular}
\end{minipage}
\endgroup
\end{minipage}
\makeatletter\def\@captype{figure}\makeatother
\caption{\textbf{Comparison with general-purpose models.} Mean reconstruction quality, anchor structure, and recorded latency on 100 clean paths, with one candidate per input. Quality means include all attempts; anchor counts use measurable outputs.}
\label{fig:llm_comparison}
\end{minipage}
\end{center}

\clearpage
\subsection{Editing study}
\label{app:user-study-materials}

Eighteen participants contributed 288 formal attempts (72 per condition) after guided practice. Figure~\ref{fig:user-study-tasks} shows four selected tasks with common targets and initial SVGs from AdaVec, LIVE, \method, or the source-SVG structural control. Participants saw anonymous method labels, with presentation order counterbalanced across the four tasks so that each condition appeared in different positions. Each attempt had a 90-second active editing limit.

\begin{center}
\begin{minipage}{\linewidth}
\centering
\includegraphics[width=0.85\linewidth]{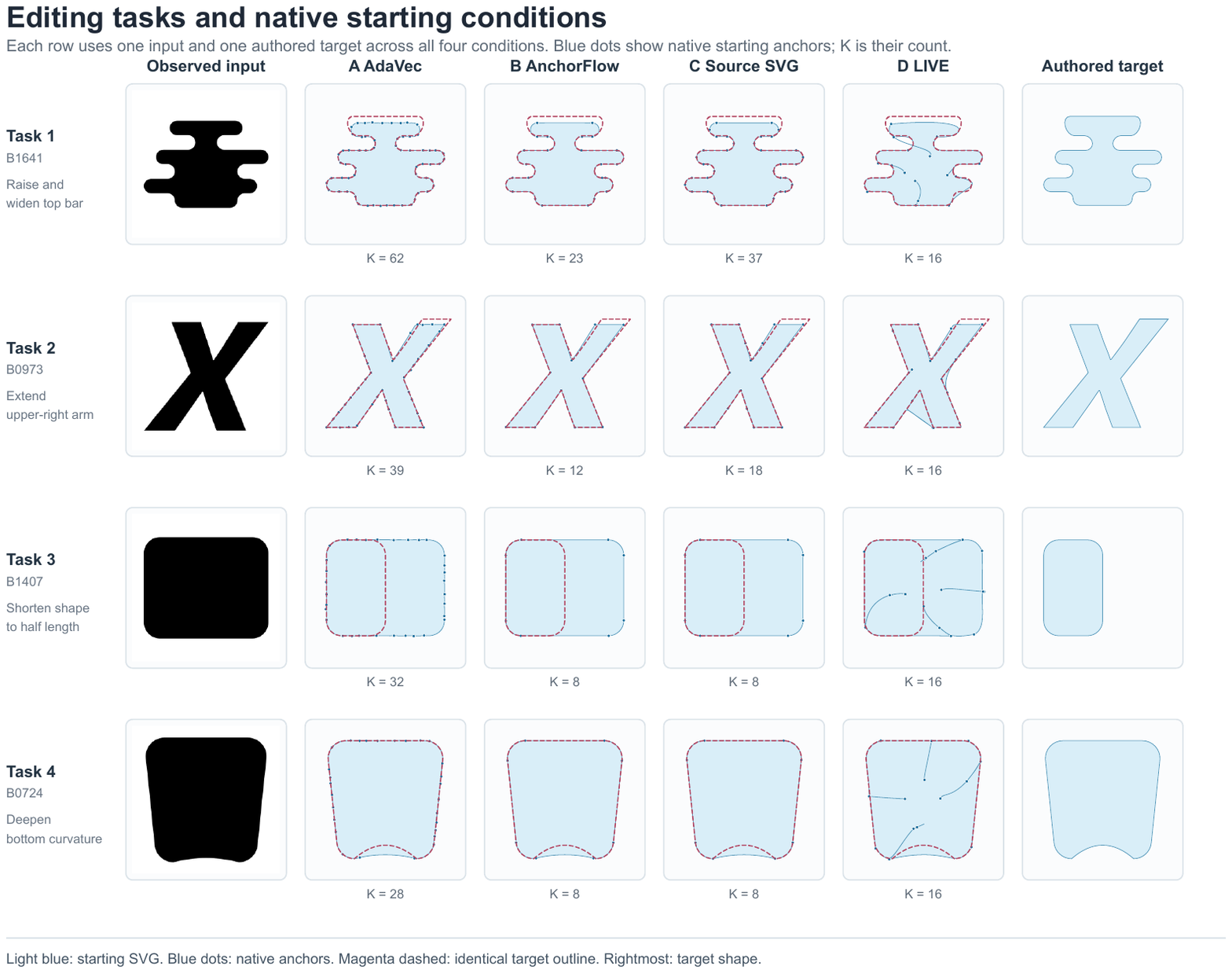}
\makeatletter\def\@captype{figure}\makeatother
\caption{\textbf{Editing tasks and starting SVGs.} Blue dots mark the $K$ initial anchors, and magenta dashed contours show targets.}
\label{fig:user-study-tasks}
\end{minipage}
\end{center}

Active time and all summaries include finished, timed-out, and abandoned attempts. Actions count insertions, deletions, and moves.

\begin{table}[!ht]
\centering
\small
\setlength{\tabcolsep}{8pt}
\renewcommand{\arraystretch}{1.18}
\caption{\textbf{Per-task editing results.} Medians over 18 attempts per cell, including all completion statuses. Top line: active time (s) / actions. Bottom line: initial $\rightarrow$ final mask IoU (\%).}
\label{tab:study-taskwise}
\begin{tabular}{@{}lcccc@{}}
\toprule
Task & AdaVec & LIVE & Source SVG & \method \\
\midrule
Top bar & \shortstack{54.0 / 16.0\\90.86 $\rightarrow$ 98.57} & \shortstack{88.5 / 20.0\\89.53 $\rightarrow$ 95.27} & \shortstack{21.0 / 6.5\\91.22 $\rightarrow$ 99.20} & \shortstack{17.5 / 3.0\\91.03 $\rightarrow$ 99.00} \\[3pt]
X arm & \shortstack{16.0 / 6.5\\94.33 $\rightarrow$ 98.18} & \shortstack{55.0 / 17.5\\87.43 $\rightarrow$ 93.20} & \shortstack{13.0 / 4.0\\94.58 $\rightarrow$ 99.18} & \shortstack{8.0 / 2.5\\94.41 $\rightarrow$ 98.76} \\[3pt]
Shorten & \shortstack{74.0 / 25.5\\49.07 $\rightarrow$ 98.19} & \shortstack{86.0 / 25.5\\49.01 $\rightarrow$ 91.21} & \shortstack{18.5 / 5.0\\49.07 $\rightarrow$ 99.17} & \shortstack{22.0 / 6.0\\49.05 $\rightarrow$ 99.12} \\[3pt]
Bottom curve & \shortstack{18.5 / 5.0\\96.59 $\rightarrow$ 99.44} & \shortstack{55.5 / 16.0\\96.51 $\rightarrow$ 98.96} & \shortstack{10.5 / 3.0\\96.72 $\rightarrow$ 99.75} & \shortstack{15.5 / 4.0\\96.67 $\rightarrow$ 99.58} \\
\bottomrule
\end{tabular}
\end{table}
\FloatBarrier

\clearpage
\subsection{Additional ablation details}
\label{app:ablation-details}

\paragraph{Readout and training controls.}
\label{app:readout-ablation}
Field-peak readout and ours share the field-pretrained AFNet; joint training from scratch learns AFNet and the anchor decoder using anchor-set supervision alone. All use the same inputs and base fitting and merging procedures. Field peaks use response floor 0.16, relative threshold $10^{-6}$, a $7\times7$ nonmaximum-suppression window and an $11\times11$ centroid window. Joint training runs for 120 epochs, matching our 60+60-epoch budget (Table~\ref{tab:network-training-settings}); its checkpoint is selected at epoch 110.

For this ablation, failed query-based reconstructions are retried at score thresholds 0.2 and 0.1, followed, if needed, by local handle repair that preserves anchor positions, order, and count. Only one joint-from-scratch output required this recovery.

\paragraph{Anchor placement at matched initial counts.}
\label{app:matched-anchor-placement}
For each of the 2,400 clean paths, all three initializers receive the count $K_i$ from our fixed-$\tau=0.2$ prediction, including the minimum-three-anchor rule and distinct contour assignment. Uniform uses equal arclength spacing. Error-greedy starts with three equally spaced anchors and repeatedly inserts the unused contour sample with largest fitting error until it reaches $K_i$. This tests placement conditional on the learned count budget. The adaptive selection used in Table~\ref{tab:shared-backend} explains its different mean initial count (11.585 versus 13.879 here).

All methods share the observed contour, coordinate system, anchor assignment, ordering, and handle fitter. Stage A fixes anchors and fits handles only, with no insertion, deletion, merging, or retry. Stage B independently restarts from the same initializations and applies common refinement, with identical insertion-budget rules, merge tolerances, and stopping conditions; method-specific retries are disabled. Budgets use the original $K_i$. Stage A yields 2,399/2,392/2,400 valid outputs for Learned/Uniform/Error-greedy; Stage B yields 2,400 for each. All inputs enter the quality means, with failures assigned zero IoU and empty anchor sets for GOSPA. Results are reported in Table~\ref{tab:matched-anchor-placement}.

\paragraph{Ours versus fixed-threshold selection.}
\label{app:anchor_selection}
Table~\ref{tab:anchor_selection} evaluates whether our anchor-selection strategy produces more compact reconstructions than a fixed threshold under the same reconstruction pipeline. Ours is compared with fixed $\tau=0.2$ using the same checkpoint, cached predictions, and fitting and merging procedures. Ours begins at $\tau=0.4$ and retries at 0.2 if reconstruction checks fail or pre-merge observed-input IoU is below 0.99. The minimum-three-anchor rule applies only at 0.2, and insertion budgets are computed per attempt.

\begin{table}[!ht]
\centering
\small
\setlength{\tabcolsep}{13pt}
\caption{\textbf{Quality and compactness under two selection policies.} Each policy produces valid outputs for all 2,400 clean and all 2,400 paired perturbed inputs. $K$ is the final anchor count.}
\label{tab:anchor_selection}
\begin{tabular}{@{}lrrrr@{}}
\toprule
 & \multicolumn{2}{c}{Clean} & \multicolumn{2}{c}{Perturbed} \\
\cmidrule(lr){2-3}\cmidrule(l){4-5}
Policy & IoU (\%) & $K$ & IoU (\%) & $K$ \\
\midrule
Fixed 0.2 & 99.583 & 10.098 & 97.696 & 13.104 \\
Ours & 99.521 & 9.533 & 97.687 & 12.550 \\
\bottomrule
\end{tabular}
\end{table}

Compared with fixed $\tau=0.2$, Ours uses 5.6\% and 4.2\% fewer final anchors on clean and perturbed inputs, respectively, while maintaining similar mean IoU.
\FloatBarrier

\clearpage
\section{Implementation and evaluation settings}
\label{app:implementation}

Experiment-specific ablation controls are given in Section~\ref{app:ablation-details}.

\subsection{Network, training, and reconstruction}
\begin{table}[!ht]
\centering
\small
\setlength{\tabcolsep}{4pt}
\renewcommand{\arraystretch}{1.12}
\caption{\textbf{Network and training configuration.} Both stages use 31,950 source paths with clean and perturbed contour views, AdamW, and effective batch size 64.}
\label{tab:network-training-settings}
\begin{tabular}{@{}p{0.23\linewidth}p{0.32\linewidth}p{0.40\linewidth}@{}}
\toprule
Setting & Field learning & Anchor decoding \\
\midrule
Trainable module & AFNet & Anchor decoder; AFNet frozen \\
Architecture & Base width 64; one input/output channel & 3 refinement layers; 4 attention heads \\
Input / features & $512\times512$ contour & $e_1$ and $d_2$ projected to 96 channels at $128\times128$, summed and normalized \\
Supervision & Contour-and-anchor field & Matched coordinates and existence at every layer \\
Epoch budget & 60 & 60 \\
Learning rate & $1.6\times10^{-4}$, then $8\times10^{-5}$ & $5\times10^{-4}$ \\
Schedule & Cosine decay & 2-epoch warmup; cosine decay \\
Selected epoch & 59 & 52 \\
\midrule
Field targets & \multicolumn{2}{p{0.74\linewidth}@{}}{Contour: amplitude 0.15, standard deviation 2 px, cutoff 4 px; anchors: standard deviation $\approx5.82$ px, floating-point centers. Targets share the input's coordinate transform.} \\
Decoder memory & \multicolumn{2}{p{0.74\linewidth}@{}}{Fine features pooled to $32\times32$, plus bottleneck tokens and positional embeddings} \\
Local sampling & \multicolumn{2}{p{0.74\linewidth}@{}}{$3\times3$ neighborhood with offsets $\{-8,0,8\}/512$} \\
Matching / losses & \multicolumn{2}{p{0.74\linewidth}@{}}{Coordinate $L_1$ matching weight 25; focal $\alpha=0.25$, $\gamma=2$; Smooth-$L_1$ transition $1/512$} \\
\bottomrule
\end{tabular}
\end{table}

Checkpoints are selected using clean and deployment validation. The validation pools contain 2,300 randomly held-out clean examples and are disjoint from both the training and test sets.

\begin{table}[!ht]
\centering
\small
\setlength{\tabcolsep}{4pt}
\renewcommand{\arraystretch}{1.12}
\caption{\textbf{Reconstruction settings.} Merge tolerances are measured relative to the path before merging, on the $256\times256$ evaluation canvas.}
\label{tab:reconstruction-settings}
\begin{tabular}{@{}p{0.24\linewidth}p{0.73\linewidth}@{}}
\toprule
Setting & Value / rule \\
\midrule
Single-path normalization & Center foreground bounding box; scale longest side to $0.72\times512$ px \\
Anchor assignment & Distinct observed-contour samples, ordered by arclength \\
Handle fitting & 128 samples per span; regularized least squares (Eq.~\eqref{eq:handle_fit}); 3 parameter-update/refitting steps; handle length capped at 1.5 times endpoint distance \\
Single-path selection & $\tau=0.4$; retry at 0.2 on fitting failure or pre-merge observed-input IoU $<0.99$; supplement to at least 3 anchors only in the 0.2 branch \\
Full-image selection & To account for possible rasterization artifacts when small components are enlarged, the accompanying code first tries $\tau=0.85$ when the component-mask area is below 1,024 pixels and the contour bounding-box span is below 64 pixels on the original canvas. Accept a valid candidate; otherwise compare candidates at $\tau=0.4$ and $0.2$. \\
Anchor insertion & Insert at high-error boundary locations and refit; budget $\min(4,\max(1,\lceil0.25K_0\rceil))$, with $K_0$ after contour assignment for that attempt \\
Merge acceptance & Preserve sampled simplicity; IoU drop $\leq0.0005$, CD increase $\leq0.01$ px, sampled arc deviation $\leq0.5$ px \\
Full-image assembly & Local handle refitting for residual self-intersections; foreground regions on white; 101 global updates, geometric-loss weight 0.1 \\
\bottomrule
\end{tabular}
\end{table}
\FloatBarrier

\clearpage
\subsection{Network architecture and curve parameterization}
\begin{figure}[!ht]
\centering
\includegraphics[width=0.88\linewidth]{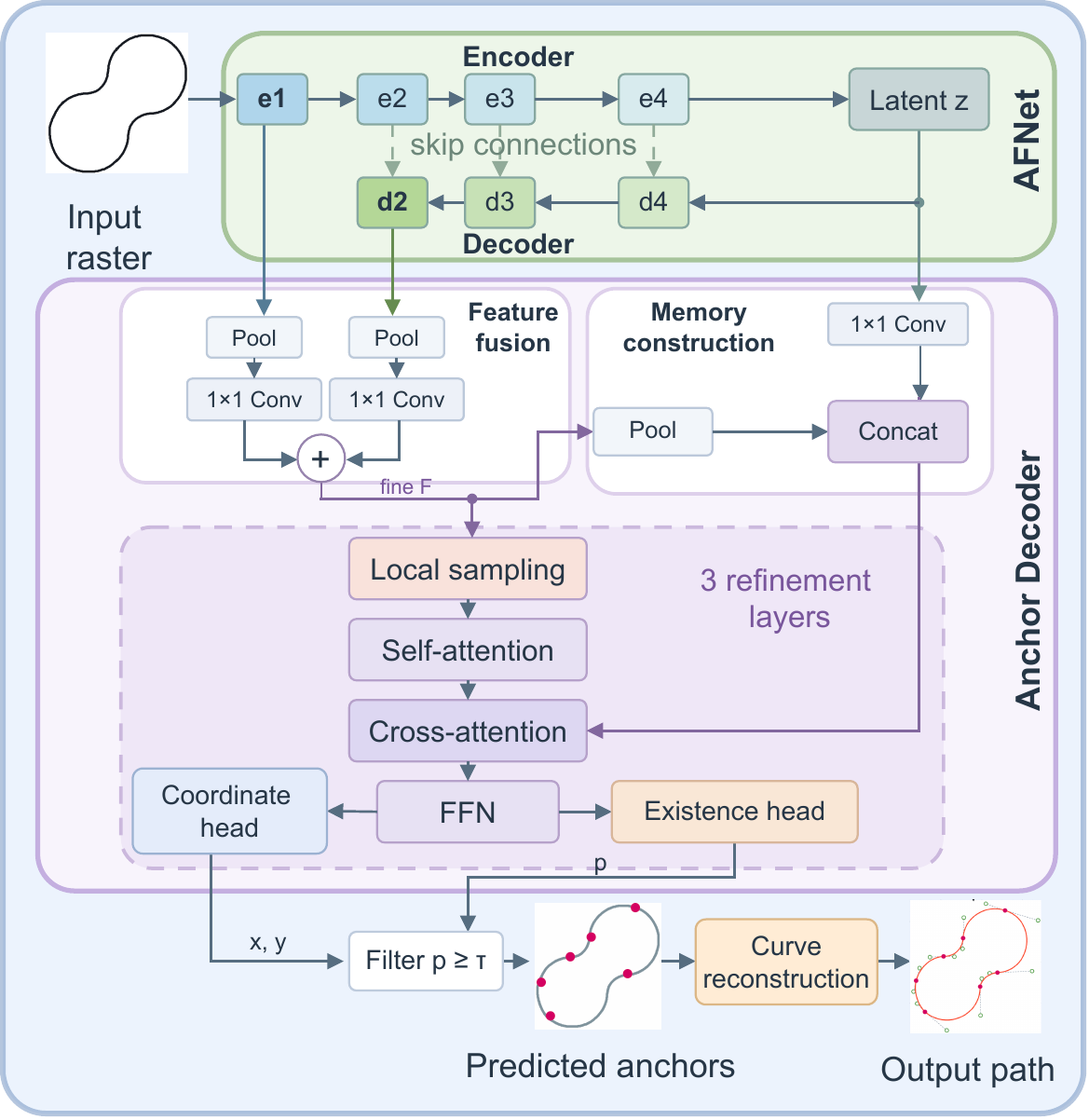}
\caption{\textbf{Network architecture.} AFNet provides encoder, decoder and bottleneck features. Pooled and projected features from $e_1$ and $d_2$ are summed to form fine features for local sampling. Pooled fine features and projected bottleneck features form the memory used by cross-attention. Three refinement layers update the anchor queries and coordinates, with an existence head predicting scores for filtering before curve reconstruction.}
\label{fig:network_details}
\end{figure}
For ordered anchors $\{a_i\}_{i=1}^{K}$, each segment is parameterized as
\begin{equation}
B_i(t;\mathbf H_i)
=
(1-t)^3a_i
+
3(1-t)^2t\,h_i^{\rm out}
+
3(1-t)t^2\,h_{i+1}^{\rm in}
+
t^3a_{i+1},
\qquad t\in[0,1],
\label{eq:bezier}
\end{equation}
where indices wrap from $K$ to $1$, and $\mathbf H_i=[h_i^{\rm out},h_{i+1}^{\rm in}]$ contains the two inner control points.
\FloatBarrier

\clearpage
\subsection{Evaluation protocol}
\label{app:protocol}

\begin{table}[!ht]
\centering
\small
\setlength{\tabcolsep}{4pt}
\renewcommand{\arraystretch}{1.16}
\caption{\textbf{Evaluation data and baseline budgets.} Full-image budgets list Noto / Fluent / HeisenVec in that order. Generative baselines produce one candidate per input.}
\label{tab:evaluation-settings}
\begin{tabular}{@{}p{0.22\linewidth}p{0.35\linewidth}p{0.38\linewidth}@{}}
\toprule
Setting & Single path & Full image \\
\midrule
Test data & 1,800 ColorSVG + 226 Bootstrap + 374 STIX paths & 100 images each from Noto Emoji, Fluent Emoji, and HeisenVec \\
Render canvas & $256\times256$, white & $512\times512$ RGB, white \\
LIVE & 16 cubic segments; 500 updates & --- \\
DiffVG & 5 paths, 20 total segments; 500 updates & 256 / 256 / 1,024 anchors; 500 updates \\
OmniSVG token cap & 1,024 & 4,096 / 4,096 / 8,192 \\
InternSVG token cap & 1,024 & --- \\
\bottomrule
\end{tabular}
\end{table}

\paragraph{Paired perturbations.}
Each source receives one fixed perturbation with raster size 192/224/256, boundary displacement 0.5--0.75 low-resolution pixels, and offset within $\pm0.2$ pixels. Scoring uses the paired clean reference.

\paragraph{Metrics.}
IoU measures foreground overlap; MSE uses values in $[0,1]$; symmetric boundary CD averages both directed nearest-neighbor distances in pixels. Anchors are segment endpoints. For LIVE and DiffVG, anchor count, Count MAE, and GOSPA include every exported path segment regardless of fill color. Count MAE and GOSPA measure agreement with source-SVG anchor layouts, which are reference segmentations rather than unique ground truth. GOSPA uses $p=2$, $\alpha=2$, and $c=0.02$ after normalization by the canvas diagonal. Full-image scoring additionally uses AlexNet v0.1 LPIPS and per-image PSNR, averaged across images.

\paragraph{Failures and coverage.}
All attempts enter quality means; complexity means use measurable outputs. 
The supplied single-path evaluator assigns missing or invalid predictions IoU $=0$, MSE $=1$, and CD $=256$. In the ablations in Tables~3--5, failed structures are empty sets for GOSPA.
Clean OmniSVG/InternSVG anchor coverage is 1,938/2,400 and 2,257/2,400. Their native-primitives structural diagnostics are excluded from best/second-best highlighting. Full-image OmniSVG anchor coverage is 95/92/85 of 100; truncated generations remain in quality scoring.

\paragraph{General-purpose models.}
Astra, Sol, Fable, Opus, and DeepSeek use their recorded high settings; Gemini uses the interface default. Measurable-anchor counts are 99/100/99/100/97/99 out of 100, respectively. All attempts enter quality summaries, and GOSPA uses canonicalized SVG anchor sets.

\paragraph{Timing scope.}
Our times include preprocessing, forward inference, fitting, and merging, excluding one-time model loading. AdaVec times use persistent-process reruns over successfully initialized paths and include point and shape optimization. Other methods retain their recorded execution scopes. Classical tracers run on CPU and learned or optimization methods on A100 GPUs.

\clearpage
\section{Additional qualitative results}
\label{app:qualitative}
These curated examples cover varied geometry and scenes; they do not estimate the frequency of particular outcomes.

\begin{figure}[!ht]
\centering
\includegraphics[width=\linewidth,height=.84\textheight,keepaspectratio]{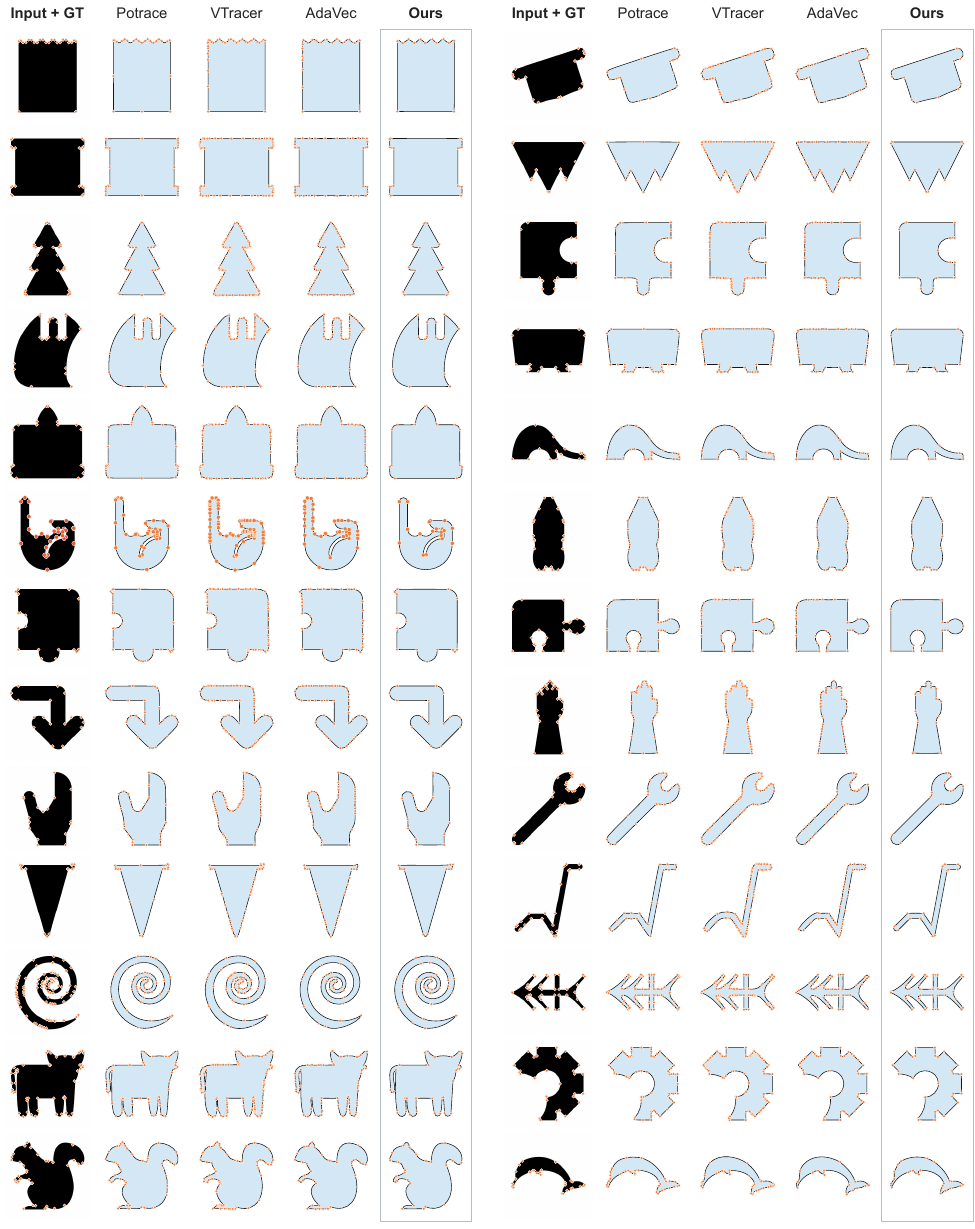}
\caption{\textbf{Clean single-path examples.} Input + GT overlays source-SVG anchors on the raster. Light-blue fills, black paths, and orange native segment endpoints are shown for all methods.}
\label{fig:supp_01_single_path}
\end{figure}

\clearpage
\begin{figure}[!ht]
\centering
\includegraphics[width=\linewidth,height=.84\textheight,keepaspectratio]{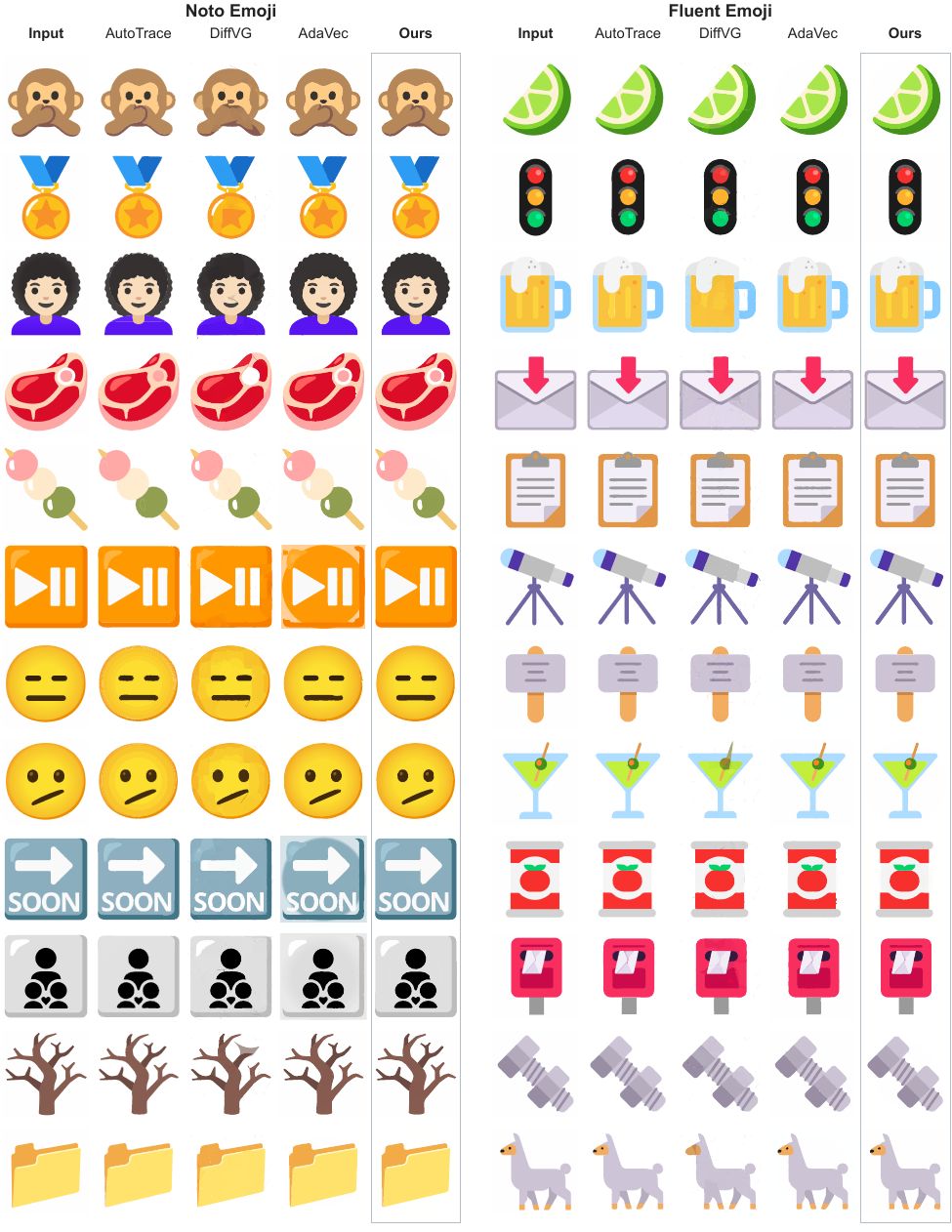}
\caption{\textbf{Noto Emoji and Fluent Emoji examples.} Native colors are retained without anchor overlays. DiffVG uses 256 anchors.}
\label{fig:supp_02_full_image_icons}
\label{fig:supp_02_fluent}
\end{figure}

\clearpage
\begin{figure}[!ht]
\centering
\includegraphics[width=\linewidth,height=.89\textheight,keepaspectratio]{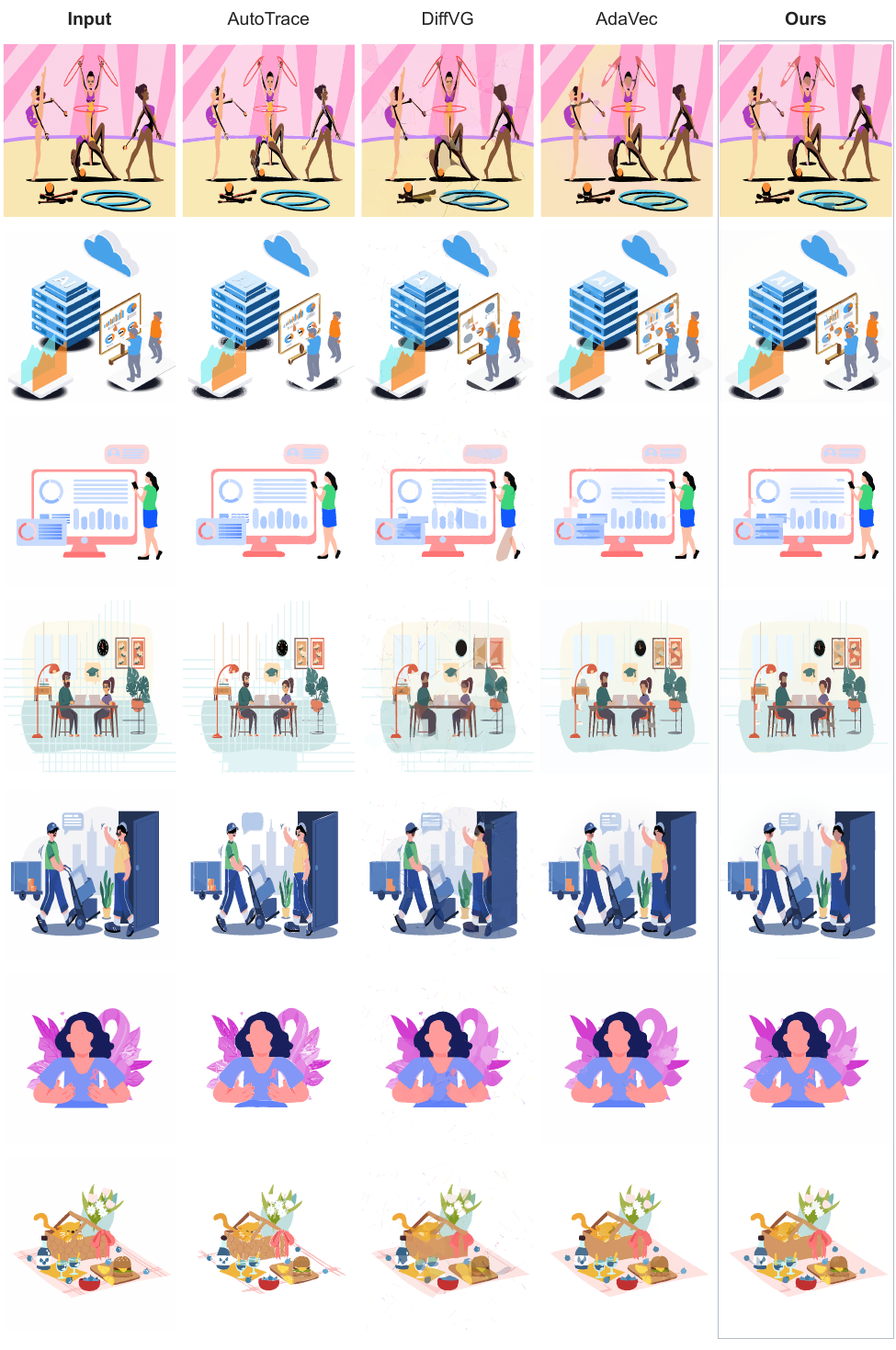}
\caption{\textbf{HeisenVec examples.} Native colors are retained without anchor overlays. DiffVG uses 1,024 anchors.}
\label{fig:supp_03_full_image_heisenvec}
\end{figure}

\clearpage
\begin{figure}[!ht]
\centering
\includegraphics[width=\linewidth,height=.89\textheight,keepaspectratio]{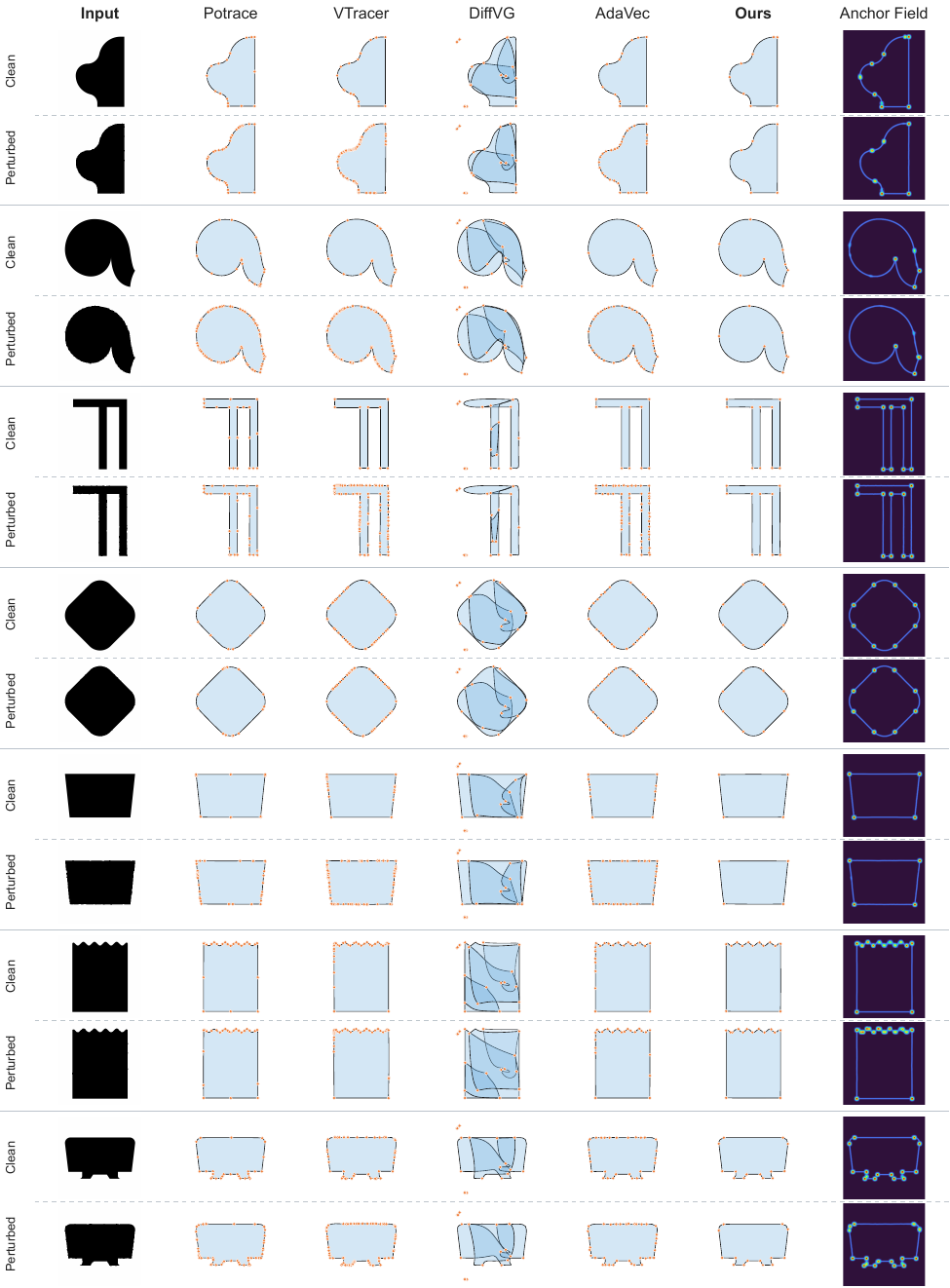}
\caption{\textbf{Paired clean and perturbed inputs.} Orange markers show native segment endpoints. Anchor-field-head responses share a fixed 0--1 color scale; the coordinate decoder reads backbone features rather than this diagnostic output map.}
\label{fig:supp_04_robustness_fields}
\end{figure}

\FloatBarrier